# Atomic-scale imaging of few-layer black phosphorus and its reconstructed edge

Yangjin Lee[1], Jun-Yeong Yoon[1], Declan Scullion[3], Jeongsu Jang[1], Elton J. G. Santos[3,4], Hu Young Jeong[2,*], and Kwanpyo Kim[1,*]

[1]Department of Physics, Ulsan National Institute of Science and Technology (UNIST), Ulsan 689-798, South Korea.

[2]UNIST Central Research Facilities (UCRF), Ulsan National Institute of Science and Technology (UNIST), Ulsan 689-798, South Korea.

[3]School of Mathematics and Physics, Queen's University Belfast, Belfast, BT95AG, United Kingdom.

[4]School of Chemistry and Chemical Engineering, Queen's University Belfast, Belfast, BT95AL, United Kingdom.

*Address correspondence to K.K. (kpkim@unist.ac.kr) or H.Y.J. (hulex@unist.ac.kr)

**Abstract**

Black phosphorus (BP) has recently emerged as an alternative two-dimensional semiconductor owing to its fascinating electronic properties such as tunable bandgap and high charge carrier mobility. The structural investigation of few-layer black phosphorus, such as identification of layer thickness and atomic-scale edge structure, is of great importance to fully understand its electronic and optical properties. Here we report atomic-scale analysis of few-layered BP performed by aberration corrected transmission electron microscopy (TEM). We establish the layer-number-dependent atomic resolution imaging of few-layer BP via TEM imaging and image simulations. The structural modification induced by the electron beam leads to revelation of crystalline edge and formation of BP nanoribbons. Atomic resolution imaging of BP clearly shows the reconstructed zigzag edge structures, which is also corroborated by van der Waals first principles calculations on the edge stability. Our study on the precise identification of BP thickness and atomic-resolution imaging of edge structures will lay the groundwork for investigation of few-layer BP, especially BP in nanostructured forms.

**1. Introduction**

A century after its discovery[1-3], black phosphorus (BP) has regained much attention as an alternative two-dimensional (2D) material owing to its promising electrical, optical, and chemical properties[4-19]. As a layered structure, BP has the largely tunable bandgap as a function of the number of layers (0.35 – 2.0 eV), which can bridge the missing bandgap range from the currently available various 2D materials[9, 15, 16, 20]. BP also poses various interesting electrical, mechanical, and optical properties, such as large tunability by strain[20-22] and high in-plane anisotropy[11, 16, 23, 24]. Especially, researchers have recently demonstrated the high charge carrier mobility from few-layer BP[5, 6, 8, 15, 16] opening up various interesting electronic applications[5, 25, 26] and fundamental studies[10, 11].

Atomic-scale structural analysis of few-layer BP is essential to fully understand its electronics and optical properties. The various defects[27, 28] have a profound effect on charge carrier dynamics, which becomes more important for the few-layer form of BP. BP nanoribbons also have various interesting properties, including edge-type-dependent electronic properties and special edge states, as shown by recent theoretical studies[22, 29, 30]. Until now, only a few experimental results on the structural characterization of BP using various microscopy techniques have been reported[9, 31-33]. Although these reports provide general structural analysis on BP, atomic-scale imaging and analysis of structural modification of BP are mainly unexplored at this point.

Here we report atomic-scale analysis of few-layered BP performed by aberration corrected transmission electron microscopy (TEM). Previously, aberration-corrected TEM imaging has been applied to various 2D materials including graphene[34-36], hexagonal boron nitride (h-BN)[37-39], and other transition metal dichalcogenides[40-42]. We establish the layer-number-dependent atomic resolution imaging of few-layer BP via TEM experiments and image simulations. In addition, we find that the electron beam can be utilized to form BP nanoribbons with crystalline edge structures. TEM imaging reveals that the BP edge shows the reconstructed edge configuration, which is also confirmed by first principles calculations with van der Waals dispersion force method. Our study on the precise identification of BP thickness and atomic resolution imaging of BP edges will lay the groundwork for investigation of electrical and optical properties of BP nanostructures.

**2. Methods**

*2.1. Black phosphorus sample preparation and thickness characterizations*

Black phosphorus (BP) crystals were purchased from Smart Elements (purity, ~99.998%). We use two different methods to prepare TEM samples. For determination of the flake thickness via independent characterization tools such as atomic force microscopy, we rely on sample preparation method 1 (See Supplementary Figure S1). We mechanically exfoliate BP crystals onto thin poly-methyl-methacrylate (PMMA)-coated $SiO_2$(300nm)/Si substrate. PMMA was spin-coated for one minute with 6000 revolutions per minute. The thin flakes were firstly identified by optical microscopy under the reflection mode. Consequently, the thickness of thin BP flakes was measured using an atomic force microscope (AFM) (DI-3100, Veeco, USA). After measurement of BP thickness (sometimes we skip the AFM imaging to avoid the possible degradation to the samples), TEM samples were fabricated using the direct transfer method[43] and PMMA layer was removed by acetone. Finally, the transmission mode in an optical microscope was used to determine the number of layers[9]. Except for AFM analysis, all processes (optical microscopy, BP exfoliation, TEM grid fabrication, and remove the polymer layer) were performed in the $N_2$ filled glove box. Another sample preparation method (sample preparation method 2) was mainly used for atomic-resolution TEM imaging. For this purpose, BP crystals were exfoliated onto $SiO_2$/Si wafers using conventional mechanical exfoliation method. Exfoliated BP flakes were transferred to Quantifoil Au TEM grids using direct transfer method[43]. Gentle plasma cleaning with $H_2$ and $O_2$ gas environment was sometimes carried out using plasma cleaner (Advanced plasma system, Gatan, USA) for 5 minutes with 10 W input power to thin BP flakes and remove surface residues. To minimize the oxidation of BP specimens, BP samples were immediately loaded into TEM chamber after sample preparation process.

*2.2. TEM characterizations and image simulations*

The atomic resolution imaging of BP was performed with a FEI Titan G2 operated at 80 kV, which is equipped with image Cs aberration corrector and monochromator. For the TEM time series, the exposure time of 0.2 seconds together with the processing time of 1.3 seconds were used. This results in the image frame time of 1.5 second per image. All the TEM image simulations were performed using MacTempas software. The crystal structure of BP with a=3.31 Å, b=10.48 Å and c=4.37Å (space group Cmca[44]) were used. The b-axis is the layer stacking direction. The normal image calculations with simulation parameters (convergence angle = 0.10 mrad, Cs = - 12 μm, mechanical vibration = 0.5 Å) in MacTempas were used. The image simulations with relevant $B_2$ coefficient of 220nm were also performed for validity check-up.

*2.3. Theoretical calculations: Van der Waals Ab Initio Calculations*

The calculations reported here are based on *ab initio* density functional theory using the SIESTA method[45] and the VASP code[46, 47]. The generalized gradient approximation[48] along with the optB88-vdW[49, 50] functional was used in both methods, together with a double-ζ polarized basis set in Siesta, and a well-converged plane-wave cutoff of 500 eV in VASP. Projected augmented wave method (PAW)[51, 52] for the latter, and norm-conserving (NC) Troullier–Martins pseudopotentials[53] for the former, have been used in the description of the bonding environment for P. NC pseudopotentials include non-linear-core corrections (NLCC)[54] to correctly account for the weak interactions between core and valence densities. The pseudocore radii $r_{NLCC}(a_o)$ (in Bohrs) together with the different $l$ channels $r_l(a_o)$ have been optimized and the values are: $r_s(a_o)$=1.83, $r_p(a_o)$=1.83, $r_d(a_o)$=1.83, $r_f(a_o)$=1.83, $r_{NLCC}(a_o)$=1.45. The shape of the NAOs was automatically determined by the algorithms described in[45]. The cutoff radii of the different orbitals were obtained using an energy shift of 50 meV, which proved to be sufficiently accurate to describe the geometries and the energetics. Atomic coordinates were allowed to relax until the forces on the ions were less than 0.04 eV/Å under the conjugate gradient algorithm. Further relaxations (0.01 eV/ Å) do not change appreciably the energetics and geometries. The lattice constants for the monolayer unit cell were optimized and found to be a=3.297 Å, b=22.1220 Å, c=4.655 Å in SIESTA which is in good agreement with the results obtained using VASP, a=3.295 Å, b=22.1219 Å, c=4.535 Å. To model the system studied in the experiments, we created large supercells containing up to 136 atoms to simulate the interface between different nanoribbon layers and edge reconstructions in the phosphorene. To avoid any interactions between supercells in the non-periodic direction, a 20 Å vacuum space was used in all calculations. In addition to this, a cutoff energy of 120 Ry was used to resolve the real-space grid used to calculate the Hartree and exchange-correlation contribution to the total energy. For the phosphorene sheets, the Brillouin zone was sampled with a 10x8x1 grid under the Monkhorst-Pack scheme[55], which gives similar results as those using a finer 17x15x1 k-sampling. In addition to this we used a Fermi-Dirac distribution with an electronic temperature of $k_BT = 20$ meV.

**3. Results and discussion**

We first study the crystal structure of BP using relatively thick (20nm or thicker) flakes. Supplementary Figure S2(a) shows a low-magnification TEM image of a typical BP flake. Without tilting samples, BP usually exhibits the crystal direction viewed along zone axis [010] as shown in Figure S2(b) and S2(d). The Fourier transform of the image clearly shows the diffraction signal with lattice parameters which are consistent with previous results[2] (Figure S2(e) and Table S1). Some transferred BP flakes display folded edge structures and this allows us to observe BP at different crystallographic directions, even without tilting of samples. Supplementary Figure S2(h) shows a TEM image around the flake edge where the crystal structure at zone axis [100] is clearly observed. At [100] zone axis, the puckered layered structure of BP can be clearly observed. The interlayer distance of BP is found to be 5.27 Å, which is consistent with previously reported results[2] (Table S1). With tilting of the specimens, atomic resolution imaging at extra zone axes is also performed as shown in Supplementary Figure S3. Especially, atomic resolution TEM images from [101] zone axis can differentiate AB and AA stacking and our observation indicates that bulk BP has mainly AB stacking (Supplementary Figure S3(f)-(j)). Our first-principles total energy calculations for different stacking also confirm that AB stacking is the most stable stacking configuration (Supplementary Figure S4).

The precise and facile identification of BP thickness is a prerequisite for investigation of various properties. We combine optical microscope (OM) imaging, atomic force microscopy (AFM), TEM imaging, and selected area electron diffraction (SAED) to precisely determine the thickness of thin flakes. The exfoliated and pre-identified BP flakes on PMMA/SiO$_2$(300nm)/Si substrate are characterized by AFM and then transferred to a Quantifoil holey TEM grid as shown in Figure 1(a)-1(c). (See Methods sections for details.) The suspended BP flakes reduce the transmittance at visible light range, which can be easily checked by transmission-mode of optical microscopy. The intensity along the dashed line in Figure 1(c) shows that the flake reduces the transmittance by about 17% and 23% compared to the region without a flake (Figure 1(d)). This shows the possibility of using the optical transmittance as a tool to determine the thickness of flakes, which will be discussed later in detail. The pre-characterized flakes are then investigated by SAED and atomic resolution TEM imaging for precise confirmation of sample thickness. Figure 1(e) shows a bright-field TEM image of the pre-identified flake. The SAED pattern is consistent with the expected crystal direction along zone axis [010] as shown in the inset of Figure 1(e). The intensity ratio ($I_{101}/I_{200}$) of diffraction patterns can be utilized for thickness determination (Figure 1(f)) similar to previous reports on 2D materials [9, 56]. By comparison with diffraction intensity simulation results (Supplementary Table S2), we confirm that the thinner area of flake has 5 layers, which is consistent with AFM result shown in Figure 1(b).

In the pre-characterized region of 5-layer thickness, atomic resolution TEM imaging and comparison with image simulation are performed. Especially, the comparison of the intensity modulation and its pattern along c-axis of BP is performed between observed TEM images and simulation results (Figure 1(g) and 1(h)). Since the phase contrast TEM image depends on the number of BP layers and defocus value, we perform a series of image simulations along [010] zone axis as a function of the layer number as well as the defocus value, where we assume the AB stacking (Supplementary Figure S5). We find that the experimentally-observed and simulated TEM image intensity pattern matches very well whereas the absolute magnitude of modulation differs by a factor of two. This factor of two discrepancy is often called as the Stobb's factor and was observed in other two dimensional materials such as graphene[34, 57, 58]. Recent studies have demonstrated that the discrepancy can be caused by the neglect of the detector modulation-transfer function (MTF) during the image simulation [59, 60]. Due to the limitation in the simulation software, we did not consider the effect of MFT for our image simulation, which can be the main reason for the observed discrepancy. After taking into account this factor, the simulation intensity pattern with defocus value of 6 nm match well with 5 layers. We confirm that the analysis is valid under imaging conditions with some residual aberrations (second-order coma $B_2$) relevant to our experiments (Supplementary Note S1, Figure S6 and Table S2).

We find that the optical transmittance of flakes is one of reliable and facile parameters for determination of flake thickness as confirmed by aforementioned complimentary characterizations as shown in Figure 2. This is of great importance since this allow us to prepare TEM samples of known thickness without AFM imaging (and resulted degradation by exposure to ambient conditions) as all the sample preparation and optical characterization process can be performed inside $N_2$-filled glovebox. In the range of 3~9 layer thickness of BP flakes we find that the optical transmittance reduces 3.3% per layer (Figure 2), which is also confirmed by SAED analysis of flakes (Supplementary Figure S7, S8, and Table S3). This is consistent with a recent report[9].

With the established procedure of atomic resolution imaging/simulation, we investigate a BP specimen where the thickness is not homogeneous in the different locations due to electron-beam-induced sputtering (Figure 3(a)). We assign the layer numbers and defocus values for the observed TEM images by comparison with simulated images through the intensity modulation and its pattern along c-axis of BP. Generally, the magnitude of intensity modulation increase as the thickness increases for a given focus value (Supplementary Figure S9 and Table S4). We can reproduce the experimentally-observed intensity patterns by simulation results with high accuracy as shown in Figure 3(b) and 3(c). The TEM images obtained from various locations of the specimen show good agreements with simulation results after the Stobb's factor correction[34, 57, 58] as shown in Figure 3(c). Through this procedure, we determine that the BP specimen imaged in Figure 3(a) has a thickness ranging from three to seven layers. This result confirms that the atomic resolution imaging alone can be utilized to reliably determine the BP thickness. One thing to note is that even and odd layer numbers produce distinct image patterns (Supplementary Figure S5). Simulations with even number of layers (for example, double layer) show the intensity modulation with the half of usual bulk lattice parameters due to the symmetric AB stacking. On the other hand, with an odd number of layers (monolayer and triple layer), the simulated images display the intensity modulation with the periodicity of usual lattice parameters of bulk BP (Supplementary Figure S5).

Now we start our discussion on the edge structure of BP. The investigation of edge structure of BP is an important topic as it significantly influences various physical properties of BP, especially for BP in nanostructured forms. There are a few recent studies on BP edges and nanoribbons mainly by theoretical calculations[29, 30]. Previously, TEM imaging has been adapted to study edge structures of various other 2D materials including graphene and hexagonal boron nitride (h-BN), which has revealed edge-specific bonding and reconstruction[35, 61-64]. On the other hand, there is only a limited number of experimental studies on BP edges by any imaging technique[31]. As shown in Supplementary Figure S10, as-prepared BP specimens exhibit amorphous edge structure, where the amorphous edge regions of several nanometers are always observed. This amorphous edge is possibly due to residual oxide layer. The plasma-treatment of a specimen can be used to reduce the amorphous edge region of the flake but this process still leaves some amorphous region (Supplementary Figure S10).

The edge structure of BP crystals can be structurally modified by e-beam irradiation during TEM imaging. Previously, the electron-beam-induced sputtering has been used for thinning down the specimen as well as cleaning[35, 37]. Remarkably, the crystalline edge structure can be obtained via this method. Figure 4(a) and 4(b) shows the changes of sample over 48-second e-beam exposure. The atoms at the amorphous edge (indicated by yellow arrows in Figure 4(a)) can be preferentially sputtered out, exposing the crystalline edge structure. The crystal direction of the exposed edge shows zigzag (ZZ) edge direction. Figure 4(c) is the zoom-in image of BP edge where the periodic edge structure over five unit-cells is clearly observed. Moreover, the image pattern at the edge shows a higher intensity modulation compared to the basal plane. This strongly suggests that there is a reconstructed edge formation.

To have a better understanding on atomic-scale structure of ZZ edge, we calculate the relaxed edge structures with various possibilities using first principles calculations with van der Waals interactions (See Methods sections). As shown in Figure 5, we find that reconstructed ZZ edges (type 1 and type 2) exhibit similar edge formation energies compared to zigzag (ZZ) or armchair (AC) edge configurations. This result is quite distinct from the graphene edge case, where the edge formation energy strongly depends on the edge type[63, 65, 66]. To compare with experimental TEM images, a series of image simulations assuming different ZZ edge structures including usual ZZ edge termination (Supplementary Figure S11) and two types of reconstructed edge configurations are undertaken (Supplementary Figure S12 and S13). Since the observed area has three-layer thickness, which is determined by the previous image pattern and intensity modulation analysis (Figure 3(b)), we focus on the simulated images from triple-layer. By comparison, we find that the observed ZZ edge is consistent with the reconstructed zigzag edge type-1 (RZZ1) edge (Figure 4(d)). The usual ZZ edge without reconstruction (Figure 4(g) and 4(h)) and RZZ2 edge (figure 4(i) and 4(j)) are not consistent with the observed image. We note that the observed RZZ1 edge was theoretically studied together with some experimental evidence but the direct atomic resolution edge imaging was not previously performed[31].

Finally, we discuss the sample thinning and BP nanoribbon formation induced by electron beam irradiation. Figures 6(a)-(e) show a time series of structural modification of BP under electron beam. The same series of images are overlaid with different colors in Figures 6(f)-(j) for easy identification of structural changes. Different colors indicate triple-layer region (blue), thicker area (pink) and amorphous regions (yellow). The sample thinning with electron-beam sputtering is observed from Figure 6(f) and 6(g); the region overlaid with pink color (thicker area) is gradually replaced by the blue region (three layers).

Electron-beam is one of useful ways to manipulate the materials at nanoscale and we demonstrate that BP nanoribbons can be formed by prolonged e-beam exposure. Figure 6(d) and 6(i) clearly show that the formation of approximately 4nm-wide BP nanoribbons with crystalline edge. After prolonged e-beam irradiation, the BP nanoribbon is amorphized to form amorphous BP nano-constriction with a less than 2 nm neck width. Consequently, the nano-constriction breaks down (See Supplementary Movie S1). The electron-beam-induced structural modification for BP seems more pronounced compared to graphene and this may be related to low energy barriers for structural phase transformation of BP[67]. The identifications of low-energy defect structures and sputtering mechanisms, such as knock-on damage and chemical etching, are important experimental issues during TEM imaging. We are currently performing the calculations of various low-energy defect structures as well as knock-on damage threshold for phosphorene. We note that a nanoribbon formation on relatively thick samples was recently reported[68].

## 4. Conclusion

In conclusion, the atomic-scale structure of few-layered BP and its reconstructed ZZ edges were investigated by Cs-corrected TEM and imaging simulation. The precise and facile characterization methods of BP thickness demonstrated in our study will lead to various fundamental studies, such as measurements of layer-number-dependent electrical and optical properties. We also demonstrate that electron beam irradiation can be used to form BP nanoribbons as well as to expose crystalline reconstructed ZZ edge for the first time. Further TEM analysis on BP is expected to shed light on various defect structures and structural degradation mechanisms.

**Acknowledgements**

This work is supported by Basic Science Research Program through the National Research Foundation of Korea (NRF) funded by the Ministry of Education (NRF-2014R1A1A2058178). H.Y.J. was supported by Creative Materials Discovery Program through the National Research Foundation of Korea (NRF) funded by the Ministry of Science, ICT and Future Planning (NRF-2016M3D1A1900035). D.S. thanks the studentship from the EPSRC-DTP award. E.J.G.S. acknowledges the use of computational resources from the UK national high performance computing service, ARCHER, for which access was obtained via the UKCP consortium and funded by EPSRC grant ref EP/K013564/1; and the Extreme Science and Engineering Discovery Environment (XSEDE), supported by NSF grants number TG-DMR120049 and TG-DMR150017. The Queen's Fellow Award through the startup grant number M8407MPH is also acknowledged.

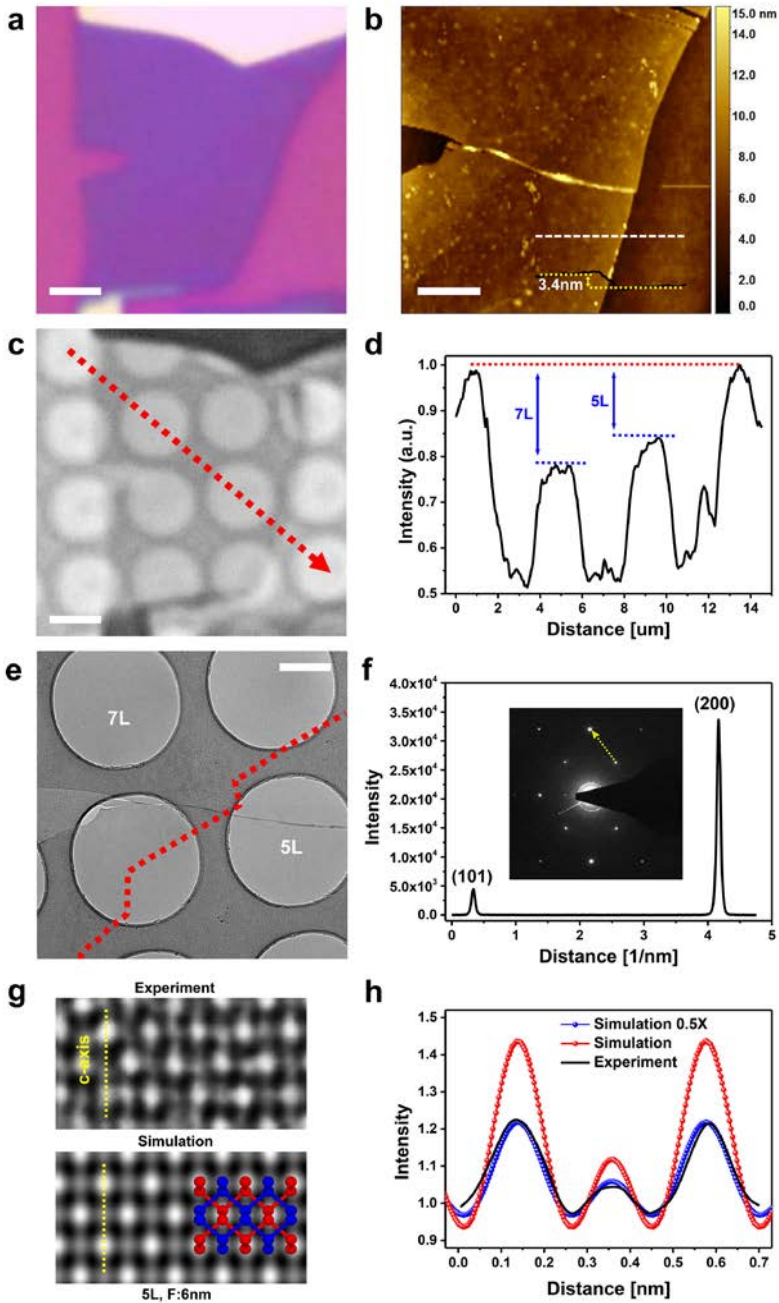

**Figure 1. Complementary thickness determination of a BP flake.** (a) Optical microscope image of a BP flake on PMMA/SiO$_2$/Si substrate. Scale bar, 2μm. (b) AFM image of the same flake. The height profile is along the white dashed line. Scale bar, 2μm (c) Transmission-mode optical microscope image of the BP flake after transfer to a TEM grid. Scale bar, 2μm (d) Intensity profile along the red arrow in panel c. (e) Bright-field TEM image of the same BP flake. Scale bar, 1μm. (f) Electron diffraction intensity profile along the yellow arrow in inset. Inset shows the diffraction pattern acquired at the 5-layer region. (g) TEM image (top) and simulated image (bottom) from 5- layer with the defocus value of 6nm. (h) Intensity profiles from TEM image (black solid line) and simulation image (red and blue solid line with circle) along the c-axis of crystal. Blue line shows the simulation profile after normalization (×0.5).

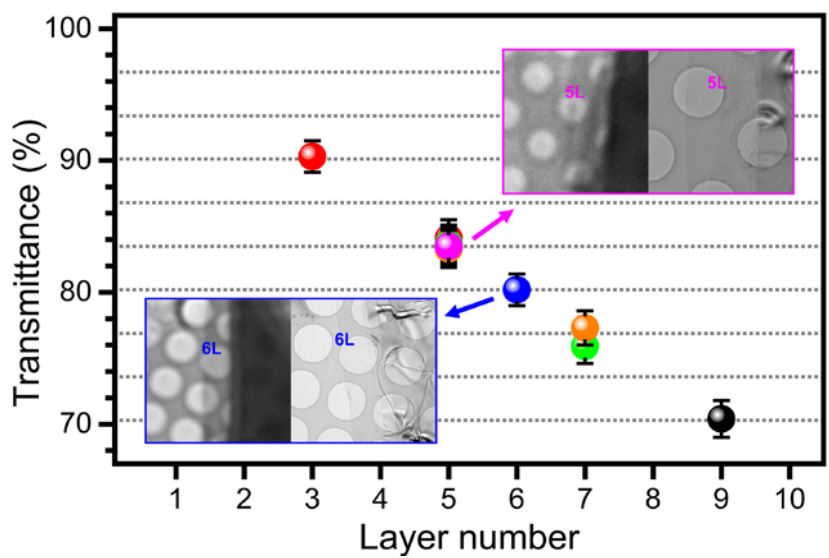

**Figure 2. Optical transmittance of suspended BP flakes as a function of layer thickness.** The observed reduction of transmittance is 3.3±0.3% per layer. Inset images show exemplary investigated samples (left: transmission-mode optical microscope image, right: TEM images of the same area).

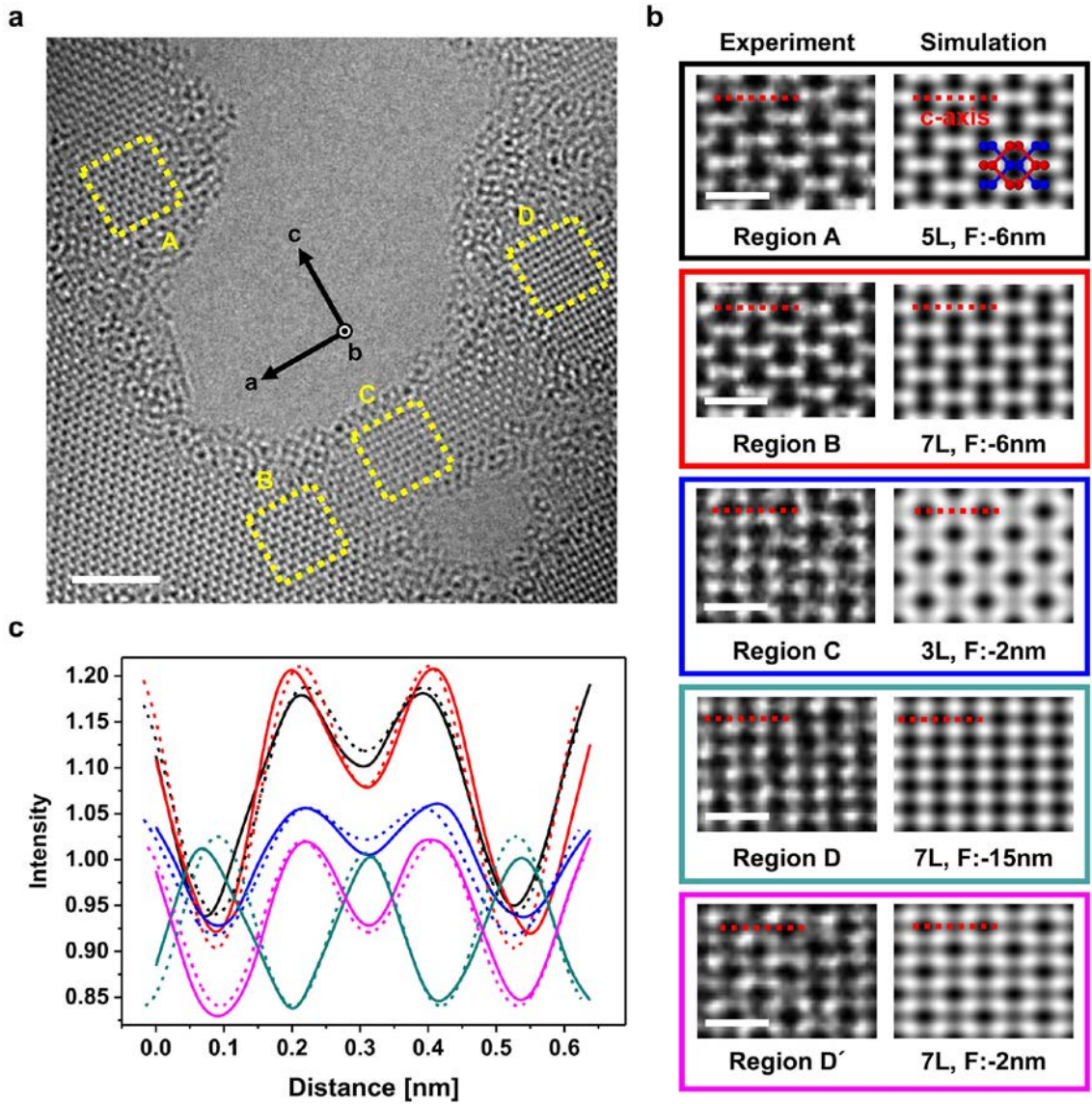

**Figure 3. Atomic resolution TEM imaging and simulation of BP.** (a) TEM image of BP showing inhomogeneous thickness at different locations. Scale bar, 2 nm. The area inside dashed squares is used for intensity profile analysis. (b) TEM images (left) acquired at different sample locations and simulated images (right) from the chosen thickness and defocus values. Scale bar, 0.5nm. (c) Intensity profiles from TEM images (solid lines) and simulation images (dashed line) along the c-axis. Black, red, blue, cyan, and pink colors indicates the data from region A, B, C, D, and D'. D' indicate the same location of D at different defocus value. All the simulation plots are normalized (×0.5). All experimental intensity profiles are the average from 5 unit cells.

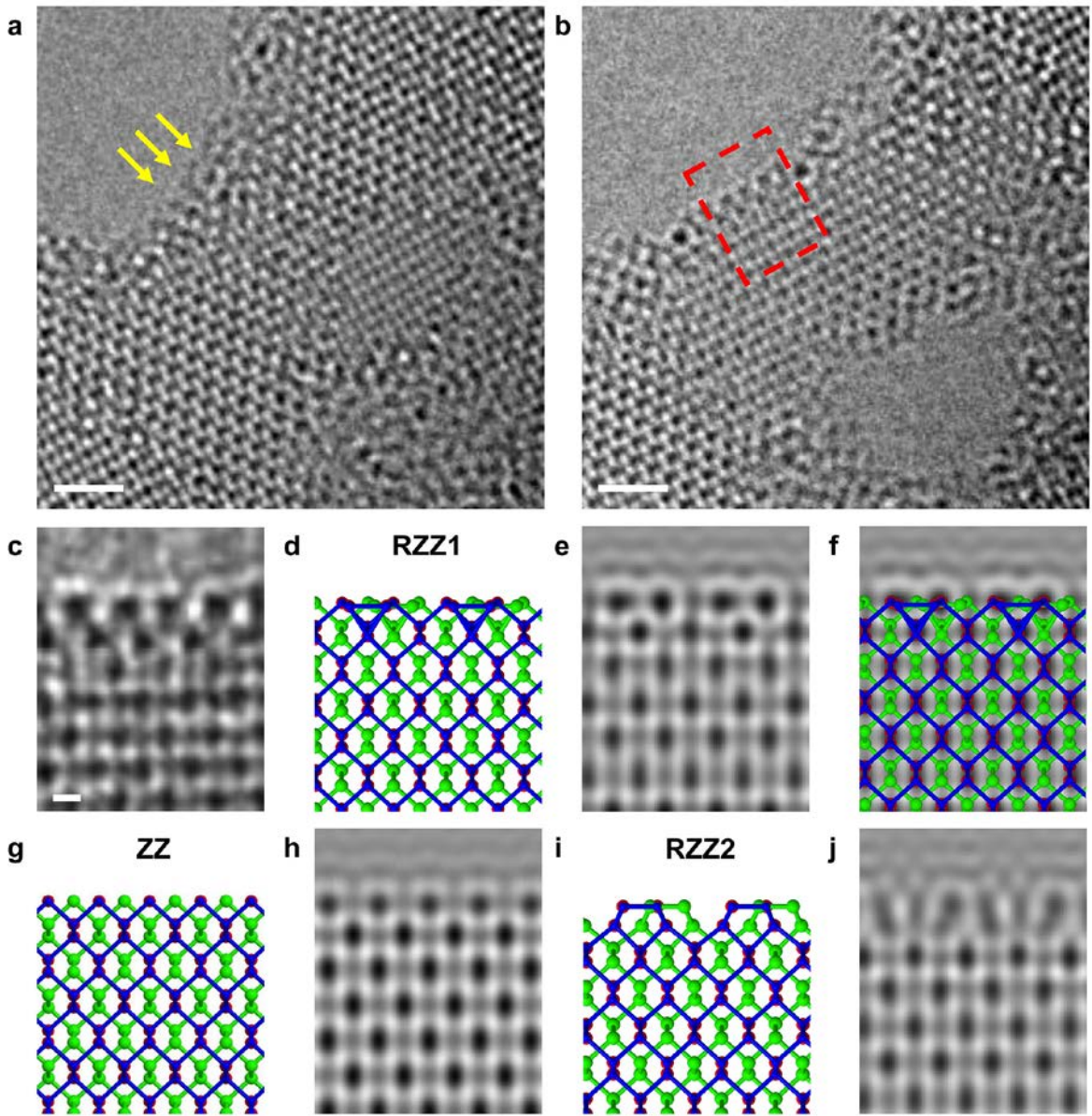

**Figure 4. Atomic resolution TEM images of BP edges.** (a) TEM image of amorphous structure at BP edge sites as indicated by yellow arrows. Scale bar, 1nm. (b) Crystalline BP edge produced by e-beam irradiation. The image was acquired after 48-second e-beam irradiation from panel a. Atoms at amorphous edge are etched by e-beam and consequently crystalline edge is exposed. The red box is the field of view for panel c. Scale bar, 1nm. (c) Zoom-in image of the BP edge. Scale bar, 0.2nm. (d) Calculated atomic model of reconstructed ZZ edge configuration 1 (RZZ1) of triple-layer BP. The atomic layer at the bottom (red) is overlapped with the top layer (blue). (e) TEM image simulation of the atomic model in panel d. (f) The same simulation image with atomic model overlay. (g) Atomic model of regular ZZ edge configuration without reconstruction. (h) TEM image simulation of regular ZZ edge structure. (i) Atomic model of reconstructed ZZ configuration 2 (RZZ2) of triple-layer BP. (j) TEM image simulation of the atomic model in panel i. Defocus value for all the simulation images is -2nm.

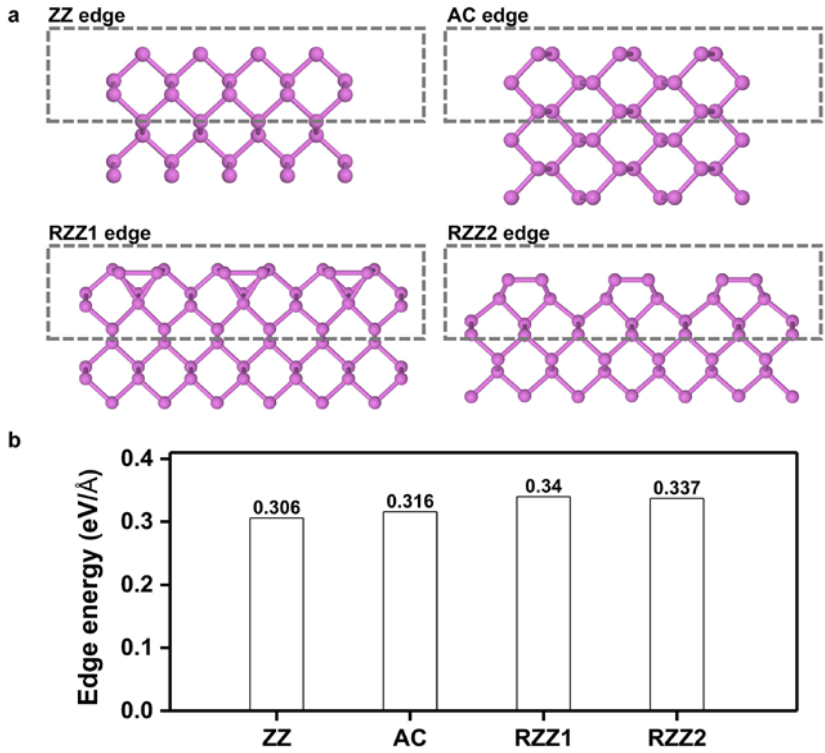

**Figure 5. BP Edge formation energy calculations.** (a) Atomistic models used for edge formation energy calculations. Pristine zigzag (ZZ), armchair (AC), reconstructed zigzag 1 (RZZ1), and reconstructed zigzag 2 (RZZ2) edge configurations are shown. (b) Edge formation energy for different edge configurations, which is given by $E_f = (1/2L)(E^{ribb} - N_P E^{bulk})$, where $E^{ribb}$ is the total energy of a ribbon with $N_P$ atoms in the supercell, and $E^{bulk}$ total energy per atom in monolayer phosphorene. $L$ is the ribbon edge length.

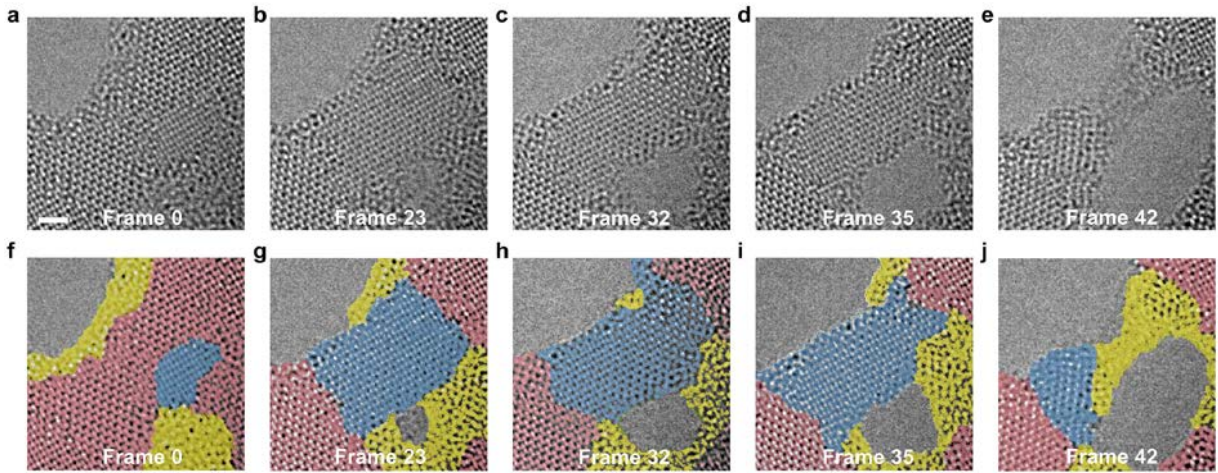

**Figure 6. Fabrication of a BP nano-constriction by electron beam irradiation.** (a-e) A time series of TEM image of BP under electron beam irradiation. Interval time of frame is 1.5 seconds. Scale bar, 1nm. (f-j) The same TEM images with color overlay. Different colors indicate triple layer (blue), thicker (pink), and amorphous (yellow) regions.